\documentstyle[aps]{revtex}

\begin{document}

\title{{\Large {\bf Improving limits on
Planck-scale Lorentz-symmetry test theories}}}

\author{{\bf Giovanni~AMELINO-CAMELIA}}
\address{Dipartimento di Fisica, Universit\`{a} di Roma ``La Sapienza''
and INFN Sez.~Roma1,\\
    P.le Moro 2, 00185 Roma, Italy}

\maketitle

\begin{abstract}\noindent%
In the recent quantum-gravity literature there has been
strong interest in the possibility
of Planck-scale departures from Lorentz symmetry.
I focus on two ``minimal'' test theories,
a pure-kinematics test theory
and an effective-field-theory-based test theory,
that could be used in this phenomenology.
Planck-scale-significant
bounds on some parameters of the two test theories can be
established using observations of TeV photons from Blazars.
Crab-nebula
synchrotron-radiation analyses, whose preliminary sensitivity estimates
had raised high hopes,
actually do not lead to any bound on the parameters of
the two test theories.
Very stringent (beyond-Planckian) limits
could be obtained, for both test theories,
if the GZK cutoff for cosmic-rays is confirmed experimentally.
\end{abstract}

\bigskip
\bigskip

\date{\today}

\maketitle

\newpage

\section{On the fate of Lorentz symmetry
in quantum spacetime}
It has been recently realized that in various
approaches to the quantum-gravity problem one encounters nonclassical
features of spacetime that lead to small departures from
Lorentz symmetry, and a quantum-gravity-motivated
phenomenology of departures from Lorentz symmetry was proposed
in Ref.~[\cite{grbgac}]. The idea
that Lorentz symmetry might be only an approximate symmetry
has then been considered
in quantum-gravity models based on spacetime foam
pictures~[\cite{garayPRL}], in loop quantum gravity
models~[\cite{gampul,mexweave}],
in noncommutative geometry
models~[\cite{gacmajid,susskind,dougnekr,gacdsr,dsrmost1}],
including some scenarios for noncommutative geometry
that are relevant in string theory~\cite{susskind,dougnekr},
and in other string-theory-inspired scenarios~[\cite{kostenew}].

At a strictly phenomenological level one can view
this interest in possible Planck-scale departures from Lorentz
symmetry as originating from the idea that the sought quantum gravity
might involve some sort of ``granularity" of spacetime (``spacetime quanta"),
and on the basis of experience with certain physical systems
(especially condensed-matter systems\footnote{Some authors have
also argued (see, {\it e.g.}, Ref.~\cite{garayPRL}) that
the quantum-spacetime environment might act in a way that
to some extent resembles the one of  a thermal environment.
It is well established (see, {\it e.g.}, Ref.~\cite{gacpi})
that in a thermal environment the energy-momentum dispersion relations
are naturally modified.})
one can expect that granularity of the medium in which propagation occurs
might lead to energy-dependent corrections~\cite{grbgac} to the dispersion
relation. At energies much larger than the particle mass but smaller
than the granularity (Plankian) energy scale,
the dispersion relation could be of the type
 \begin{equation}
 m^2 \simeq E^2 - \vec{p}^2
 +  \eta \vec{p}^2 \left({E^n \over E^n_{p}}\right)
 + O({E^{n+3} \over E^{n+1}_{p}})
 ~
 \label{displead}
 \end{equation}
where $E_p \simeq 1.2 {\cdot} 10^{16} TeV$ is the Planck scale,
$\eta$ parametrizes the ratio between the Planck scale and
the scale of quantization of spacetime,
and the power $n$ is a key characteristic of the magnitude of the effects
to be expected.

The literature on the subject is characterized primarily by a debate
on weather one should or should not assume that these departures
from Lorentz symmetry can be introduced within
the framework of effective
low-energy field theory.
Those who are most concerned about the reliability of this assumption
opt to limit the phenomenology to contexts in which one is
able to perform a pure-kinematics test.
Those who assume the validity of
low-energy effective field theory can use it to describe some aspects of
dynamics and therefore have available a wider class
limit-setting opportunities.

For these two perspectives on the problem there are correspondingly
two ``natural''~[\cite{newlim}] starting points as test theories
on which to base the phenomenology.
When one does not assume the validity of
low-energy effective field theory,
it is natural to set up a pure-kinematics
test theory based on the dispersion relation (\ref{displead})
and on the assumption that the energy-momentum conservation
rules are not Planck-scale modified (even though there are
frameworks in which instead
both the dispersion relation and the rules of energy-momentum
conservation would be modified; see, {\it e.g.}, Ref.~[\cite{gacdsr}]).
The pure-kinematics objectives are also fully compatible with
the assumption of ``universality'', {\it i.e.} the modification
of the dispersion relation is assumed to affect in the same
way (same values of parameters) all particles.
These hypotheses constitute the ``minimal AEMNS test theory''~[\cite{newlim}]
(where ``AEMNS'' refers to the authors of Ref.~[\cite{grbgac}],
which introduced the first building blocks of this test theory).
While it is rather natural to get started on this pure-kinematics
phenomenology using the ``minimal AEMNS test theory'', of course
one can contemplate various generalizations (``non-minimal versions
of the AEMNS test theory''), including the one of a ``non-universal''
Planck-scale modification of the dispersion relation.

When adopting the alternative perspective, which assumes the
applicability of effective low-energy field theory,
one can again take the dispersion-relation parametrization of
Eq.~(\ref{displead}), but consistency with the use of
the field-theoretic setup imposes to renounce immediately
to the simplifying hypothesis of universality.
In fact, the constraint of introducing the new effects
within the field-theoretic formalism, with its reference to
a Lagrangian density, restricts the types of modifications one can consider,
and in particular it is easy to show~[\cite{gampul,rob}]
that the allowed terms in the Lagrangian density lead to
a polarization dependence of the effect for photons:
in the field-theoretic setup it turns
out that when right-circular polarized photons satisfy the
dispersion relation $E^2 \simeq p^2 + \eta_\gamma p^3/E_p$ then necessarily
left-circular polarized photons satisfy the ``opposite sign"
dispersion relation $E^2 \simeq p^2 - \eta_\gamma p^3/E_p$.
For spin-$1/2$ particles the analysis reported in
Ref.~[\cite{rob}] leads to the introduction
of  two independent dispersion-relation-deformation
parameters, one for each helicity.
Since photons experience a complete correlation of the
sign of the effect with polarization
it appears natural to assume (at least in the first works
exploring this scenario)
that also for fermions
the modification of the dispersion relation should have the same
magnitude for both signs of the helicity, but have a correlation
between the sign of the helicity and the sign of the dispersion-relation
modification.
This would correspond to the natural-looking
assumption that the Planck-scale
effects are such that in a beam composed of randomly selected particles
the average speed in the beam is still governed by ordinary special
relativity (the Planck scale effects average out summing over
polarization/helicity).
These observations provide the ingredients of the ``minimal GPMP test
theory"~\cite{newlim} (``GPMP''
from the initials of the authors of Refs.~\cite{gampul,rob},
where most of the ingredients of this scenario were introduced).

In these notes I will comment on certain types of data that
are being considered as opportunities to test scenarios for
Planck-scale violations of Lorentz symmetry, and analyze their
applicability to the two ``minimal'' test theories.
My discussion is not of the type ``status of experimental limits
on the test theories'', but rather I intend to illustrate how
the (apparently small) differences between the two minimal test theories
can affect significantly the phenomenology.
I will therefore not consider all the opportunities for testing that
have been discussed in the literature. I will just
focus on a few illustrative examples.

\section{Derivation of limits from time-of-flight analyses}
The most popular strategy for establishing experimental limits
on Planck-scale modifications of the dispersion relation
is based~\cite{grbgac}
on the fact that both in
the minimal AEMNS test theory and in the
minimal GPMP test theory
one expects a wavelength dependence of the speed of photons,
by combining the modified dispersion relation and
the relation $v = dE/dp$. At ``intermediate energies" ($m < E \ll E_p$)
this velocity law will take the form
\begin{equation}
v \simeq 1 - \frac{m^2}{2 E^2} +  \eta \frac{n+1}{2} \frac{E^n}{E_p^n}
~.
\label{velLIVbis}
\end{equation}
Whereas in ordinary special relativity two photons ($m=0$)
emitted simultaneously would always reach simultaneously a far-away detector,
according to (\ref{velLIVbis}) two simultaneously-emitted
photons should reach the detector at different times
if they carry different energy. Moreover, in the case of the GPMP test
theory even photons with the same energy would arrive at different
times if they carry different polarization.
In fact, while the minimal AEMNS test theory assumes universality,
and therefore
a formula of this type would apply to photons of any polarization,
in the GPMP test theory, as mentioned, the sign of the effect
is correlated with polarization.
As a result, while the AEMNS test theory is best tested by
comparing the arrival times of particles of different energies,
the GPMP test theory is best tested by considering the highest-energy
photons available in the data and looking for a sizeable spread
in times of arrivals (which one would then attribute to
the different speeds of the two polarizations).

This time-of-arrival-difference effect
can be significant\cite{grbgac,billetal}
in the analysis of short-duration bursts of photons that reach
us from far away sources.

In the near future an excellent opportunity to test this
effect will be provided by observations of gamma-ray bursters.
For a gamma-ray burst it is not uncommon that the time travelled
before reaching our Earth detectors be of order $T \sim 10^{17} s$.
And microbursts within a burst can have very short duration,
as short as $10^{-3} s$ (or even $10^{-4} s$), and this
means that the photons
that compose such a microburst are all emitted at the same time,
up to an uncertainty of $10^{-3} s$.
Some of the photons in these bursts
have energies that extend at least up to the $GeV$ range.
For two photons with energy difference of order $\Delta E \sim 1 GeV$
a $\eta \Delta E/E_p$ speed difference over a time of travel
of $10^{17} s$
would lead to a difference in times of arrival of
order
\begin{equation}
\Delta t \sim \eta T \Delta \frac{E}{E_p} \sim 10^{-2} s
~,
\label{tdelay}
\end{equation}
which is significant (the time-of-arrival differences would be larger
than the time-of-emission differences within a single microburst).

For the minimal AEMNS test theory the
Planck-scale-induced time-of-arrival difference
could be revealed\cite{grbgac,billetal}
upon comparison of the ``average arrival time''
 of the gamma-ray-burst signal (or better a microburst within the
 burst)
in different energy channels.
The GPMP test theory would be most effectively tested by looking
for a dependence of the time-spread of the microbursts that grows
with energy.

Since the quality of relevant gamma-ray-burst data
is still relatively poor,
the present best limit was obtained
in Ref.~\cite{billetal} using a slightly different type of observations:
the negative results
of a search of time-of-arrival/energy correlations
for a TeV-gamma-ray short-duration flare from the
Markarian 421 blazar
allowed to deduce the limit $|\eta| < 3 {\cdot} 10^{2}$.
For the minimal GPMP test theory one also correspondingly
concludes that $|\eta_\gamma|$ is of $O(10^{2})$ or smaller.

Considering the achievable sensitivities
it appears~\cite{glast} that
the next generation of gamma-ray telescopes,
such as GLAST~\cite{glast},
might be able to test very significantly (\ref{velLIVbis})
in the case $n=1$, by possibly pushing the
limit on $\eta$ far below $1$.
In order to achieve this level of sensitivity it might however
be necessary to gain some understanding of certain types
of potentially troublesome at-the-source effects,
which were discussed in Ref.~\cite{piranKARP}.

\section{Limits obtained from observed absorption
of TeV photons from Blazars}
In addition to
the possible manifestation in time-of-arrival/energy correlations,
the quantum-gravity-scale modifications of the dispersion relation
could have~\cite{kifu,ita,aus,gactp}
observably-large implications for what concerns the opacity
of our Universe to various types of high-energy particles.
Of particular interest
is the fact that, according to the conventional (classical-spacetime)
description,
the infrared diffuse extragalactic background
should give rise to strong absorption of ``$TeV$ photons"
(here understood as photons with energy $1 TeV < E < 30 TeV$).
The relevant process is of course $\gamma \gamma \rightarrow e^+ e^-$.
With a given dispersion relation and a given rule for energy-momentum
conservation one has a complete ``kinematic scheme" for the analysis
of the requirements
for particle production in collisions (or decay processes).
Both the minimal AEMNS test theory and the minimal
GPMP test theory involve
modified dispersion relations and unmodified laws of energy-momentum
conservation (the fact that the law of energy-momentum conservation
is not modified is explicitly among the ingredients of the AEMNS
test theory, while in the GPMP test theory it follows from
the adoption of low-energy effective field theory).

Combining a modified dispersion relation
with unmodified laws of
energy-momentum conservation
one naturally finds a modification
of the threshold requirements for
the  $\gamma \gamma \rightarrow e^+ e^-$
process.
Let us in particular consider the dispersion relation (\ref{displead}),
with $n=1$, in the analysis of a
collision between
a soft photon of energy $\epsilon$
and a high-energy photon of energy $E$.
For given soft-photon energy $\epsilon$,
the process $\gamma \gamma \rightarrow e^+ e^-$
is allowed only if $E$ is greater than a certain
threshold energy $E_{th}$ which depends on $\epsilon$ and $m_e^2$.
For $n=1$, combining (\ref{displead}) with unmodified
energy-momentum conservation,
this threshold energy
(assuming $\epsilon \ll m_e \ll E_{th} \ll E_p$)
is estimated as
\begin{equation}
E_{th} \epsilon + \eta \frac{E_{th}^3}{8 E_p} \simeq m_e^2
~.
\label{thrTRE}
\end{equation}
The special-relativistic result $E_{th} = m_e^2 /\epsilon$
corresponds of course to the $\eta \rightarrow 0$ limit
of (\ref{thrTRE}).
For $|\eta | \sim 1$ the Planck-scale correction can be
safely neglected as long as $\epsilon > (m_e^4/E_p)^{1/3}$.
But eventually, for sufficiently small values of $\epsilon$
(and correspondingly large values of $E_{th}$) the
Planck-scale correction cannot be ignored.

In particular, if the photon of energy $\epsilon$ is part of the
infrared diffuse extragalactic background
and the photon  emitted by
a blazar is of TeV-range energy
one finds that the prediction for
absorption of the hard photon by the
infrared diffuse extragalactic background
is significantly modified.

The fact that
the observations still give us only a preliminary picture of absorption
together with the fact that there is a significant level of
uncertainty in phenomenological models of $TeV$ blazars
and in phenomenological models of the
density of the infrared diffuse extragalactic background
does not allow us to convert these observations into tight limits
on departures from the classical-spacetime analysis.
However, even just the basic fact
that we see absorption of $TeV$ $\gamma$-rays, as now suggested
by several analyses,
allows to derive~\cite{newlim} the limit $\eta \geq - 46$
({\it i.e.} either $\eta$ is positive or $\eta$ is negative with
absolute value smaller than 46).

Up to this point in this section my discussion
is strictly applicable only to the minimal AEMNS test theory.
For the case of the minimal GPMP test theory
the analysis must be modified to take into account the fact
that the modification of the dispersion relation carries
opposite sign for the two polarizations of the photon
(and for the two helicities of the electron/positron).
In light of this polarization dependence in the
minimal GPMP test theory
only one of the two polarizations of the photon
could escape absorption.
Whereas in the classical-spacetime picture one would expect
a cutoff behaviour to affect the entire spectrum
(and in the minimal AEMNS test theory one might find no
cutoff at all), in the minimal GPMP test theory one would
expect a cutoff behaviour only for a part of the spectrum,
while the rest could be unaffected by the cutoff.

\section{Derivation of limits from analysis of synchrotron radiation}
A recent series of
papers\cite{jacoNATv1,newlim,jaconature,tedreply,carrosync,nycksync,tedsteck}
has focused on the possibility to set limits on Planck-scale modified
dispersion relations focusing on their implications for synchrotron radiation.
In Ref.\cite{jacoNATv1} the starting point is the observation
that in the conventional (Lorentz-invariant) description of synchrotron
radiation one can estimate the characteristic energy $E_c$ of
the radiation through a heuristic analysis~\cite{jackson}
leading to the formula
\begin{equation}
E_c \simeq {1 \over
R {\cdot} \delta {\cdot} [v_\gamma - v_e]}
~,
\label{omegacjack}
\end{equation}
where $v_e$ is the speed of the electron,
$v_\gamma$ is the speed
of the photon, $\delta$ is the emission angle
for the radiation, and $R$ is the radius of curvature of
the trajectory of the electron.

Assuming that the only Planck-scale modification in this formula
should come from the velocity law (described using $v=dE/dp$
in terms of the modified dispersion relation),
one finds that in some instances the characteristic energy of
synchrotron
radiation may be significantly modified by the presence of
Planck-scale departures from Lorentz symmetry.
As an opportunity to test such a modification of the
value of the synchrotron-radiation characteristic energy one
can hope to use some relevant data~\cite{jacoNATv1,jaconature}
on photons detected from the Crab nebula.
This must be done with caution since
the observational information on synchrotron radiation being emitted
by the Crab nebula is rather indirect: some of the photons we observe
from the Crab nebula are attributed to synchrotron processes on the basis
of a promising conjecture, and the value of the
relevant magnetic fields is not directly measured.

Assuming that indeed the observational situation has been properly
interpreted, and relying on the mentioned assumption that
the only modification to be taken into account is the
one of the velocity law,
this type of analysis has the potential to establish
very stringent limits on some Lorentz-symmetry-breaking parameters.
However this will of course depend on the detailed structure of
the Lorentz-symmetry-breaking scheme.
In particular, it turns out that the minimal test theories that I
am considering cannot be constrained in this way.

For what concerns the minimal AEMNS test theory
it is important to realize that
synchrotron radiation is due to the acceleration
of the relevant electrons and therefore implicit
in the derivation of the formula (\ref{omegacjack})
is a subtle role for dynamics~\cite{newlim}.
From a field-theory perspective the process of
synchrotron-radiation emission
can be described in terms
of Compton scattering of the electrons
with the virtual photons of the magnetic field.
One would therefore
be looking deep into the dynamical features of the theory.

The minimal AEMNS test theory does assume a
modified dispersion relation
of the type (\ref{displead}) universally applied
to all particles,
but it is a pure-kinematics framework and, since the analysis
involves some aspects of dynamics,
it cannot be tested using a Crab-nebula
synchrotron-radiation analysis.

The GPMP test theory relies on a description of dynamics
within the framework of effective low-energy theory,
but, as mentioned, this in turn ends up implying that it is not
possible to assume
that a dispersion relation
of the type (\ref{displead}) universally applies
to all particles.
Actually the two polarizations of photons must, within
this framework, satisfy different (opposite-sign Planck-scale
corrections) dispersion relations. And for the description
of electrons one naturally encounters two more free parameters,
which in my ``minimal GPMP test theory'' are also of equal magnitude
and opposite sign (in order to preserve $c$ as the ``average speed'',
as discussed in Section~1).
As a result the ``minimal GPMP test theory"
automatically evades the type of constraint that could come
from the Crab-nebula
synchrotron-radiation analysis: since the two helicities are
affected by opposite-sign modifications of the dispersion relation,
at least one of them must be a positive-sign-type modification.

\section{Derivation of limits from analysis of UHE cosmic rays}\label{uhecr}
In Section~3 I discussed the implications of possible
Planck-scale effects
for the process $\gamma \gamma \rightarrow e^+ e^-$,
but of course this is not the only process in which Planck-scale
effects can be important. In particular, there has been strong
interest\cite{kifu,ita,aus,gactp,tedtwo,gacpion,orfeupion,nguhecr}
in ``photopion production", $p \gamma \rightarrow p \pi$,
where again the combination of
(\ref{displead}) with unmodified
energy-momentum conservation
leads to a modification of the minimum proton energy required
by the process (for given photon energy).
In the case in which the photon energy is the one typical of CMBR photons
one finds that the threshold proton energy can be significantly shifted
upward (for negative $\eta$), and this
in turn should affect at an observably large level the
expected ``GZK cutoff" for the observed cosmic-ray spectrum.
Observations reported by the AGASA\cite{agasa} cosmic-ray
observatory provide some encouragement for the idea of
such an upward shift of the GZK cutoff, but the issue
must be further explored\footnote{This AGASA-data-based ``GZK puzzle''
has been very important in providing motivation for studies
of Planck-scale departures from Lorentz symmetry. Even if a future
improved understanding of the cosmic-ray spectrum ends up removing
the puzzle, the lessons learned for the study of the quantum-gravity
problem will still be very valuable. An analogous situation has
been rather recently encountered in the particle-physics literature:
the discussion of the so-called ``centauro events'' led to strong
theoretical progress in the understanding of the possibility
of ``misaligned vacua'' in QCD (see, {\it e.g.}, Ref.~\cite{bjpap}),
and this progress on the theory side
remains valuable event though now
most authors believe that ``centauro events''
might have been a mirage }.
Forthcoming cosmic-ray observatories, such as Auger\cite{auger},
should be able\cite{kifu,gactp} to fully investigate this possibility.

In this context the comparison of the AEMNS test theory and the GPMP
test theory is rather straightforward. We are in fact considering a
purely kinematical effect: the shift of a threshold requirement.
For the minimal AEMNS test theory there is a clear prediction that
for negative $\eta$ there should be an upward shift
of the GZK threshold.
For the minimal
GPMP test theory,
where for one of the helicities of the proton the dispersion relation
is of negative-$\eta$ type and for the other
helicity the dispersion relation
is of positive-$\eta$ type,
one would expect roughly one half of the UHE protons to evade
the GZK cutoff, so the cutoff would still be violated but in
a softer way than in the case of the AEMNS test theory with
negative $\eta$.

If the Auger data should actually show evidence of the
expected GZK cutoff, then the case of negative $\eta$ for the minimal
AEMNS test theory would be severely constrained (both for $n=1$
and $n=2$),
and the fermion-sector parameter
of the minimal GPMP test theory would also be severely constrained.
Indeed, in the minimal AEMNS test theory
violations of the GZK cutoff are predicted for negative $\eta$
(while they are not present in the positive-$\eta$ case),
while in the minimal GPMP test theory violations of
the GZK cutoff (although less numerous than expected in the
minimal AEMNS test theory with negative $\eta$) are always expected,
independently of the sign of the fermion-sector parameter (depending on the
sign of fermion-sector parameter
the protons that violate the GZK cutoff would
have a corresponding helicity).

\section{Closing remarks}
I focused on the minimal  GPMP test theory and
minimal AEMNS test theory, representing respectively the field-theory
intuition and the no-field-theory intuition in the study
of Planck-scale departures from Lorentz symmetry.
I found that the differences between these two test theories,
even though they might at first appear to be rather marginal
differences
(both test theories essentially adopt the same type of modification
of the dispersion relation),
lead to significant differences in the outcome of certain
phenomenological analyses.
This should be kept in mind in the relevant ``quantum-gravity
phenomenology''~\cite{polonpap} literature.
There have been several papers claiming to improve limits
on Planck-scale modifications
of the dispersion relation, but the studies did not
rely on a well-defined test theory.
From outside the quantum-gravity-phenomenology community
these papers were perceived as a gradual improvement
in the experimental bounds on the overall idea
of Planck-scale departures from Lorentz symmetry,
to the point that there is now a wide-spread perception
that in general departures from Lorentz symmetry are already
experimentally constrained to be far beyond the Planck-scale.
Instead, as shown by the analysis of the two ``minimal'' test
theories I considered,
some of the experimental-limit opportunities that have generated
most excitement in the recent literature
are inapplicable to some meaningful scenarios for Planck-scale
departures from Lorentz symmetry.

For the objectives I was pursuing here it was necessary to focus
indeed on rather similar test theories, so that I could illustrate
the fact that even
relatively small differences in the structure of the test
theories can affect significantly the phenomenology.
Both the minimal AEMNS test theory and the minimal GPMP test theory
adopt the same type of dispersion relation and both assume that
Planck-scale effects would break Lorentz symmetry.
Even more significant differences, from the perspective of phenomenological
analyses, should be expected in test theories that explore the
possibility that the Planck-scale would ``deform'' Lorentz symmetry
(in the sense of the ``doubly-special relativity''
scenario~\cite{gacdsr,dsrmost1,dsrmost2}).
Work on the development of such a doubly-special-relativity
test theory is in progress
(see, {\it e.g.}, Ref.~\cite{dsrphen}), but the analysis
is still at too early a stage for me to comment on it here.

%

%
%
%
%
%
%
%
%



\begin{thebibliography}{}

\bibitem{grbgac} G.~Amelino-Camelia, J.~Ellis, N.E.~Mavromatos
and D.V.~Nanopoulos:
hep-th/9605211,
Int.~J.~Mod.~Phys.~A \textbf{12}, 607 (1997);
G.~Amelino-Camelia, J.~Ellis, N.E.~Mavromatos,
D.V.~Nanopoulos and S.~Sarkar:
astro-ph/9712103,
{\ Nature} \textbf{393}, 763 (1998).

\bibitem{garayPRL} L.J.~Garay:
{\ Phys.~Rev.~Lett.}~\textbf{80}, 2508 (1998).

\bibitem{gampul} R.~Gambini and J.~Pullin:
{\ Phys.~Rev.}~D \textbf{59}, 124021 (1999).

\bibitem{mexweave} J.~Alfaro, H.A.~Morales-Tecotl and L.F.~Urrutia:
{\ Phys.~Rev.~Lett.}~\textbf{84}, 2318 (2000).

\bibitem{gacmajid} G.~Amelino-Camelia and  S.~Majid:
{\ Int.~J.~Mod.~Phys.}~A \textbf{15}, 4301 (2000).

\bibitem{susskind} A.~Matusis, L.~Susskind and N.~Toumbas:
{\ JHEP} \textbf{0012}, 002 (2000).

\bibitem{dougnekr} N.R.~Douglas and N.A.~Nekrasov:
Rev.~Mod.~Phys.~\textbf{73}, 977 (2001).

\bibitem{gacdsr} G.~Amelino-Camelia: gr-qc/0012051,
{\ Int.~J.~Mod.~Phys.}~D \textbf{11}, 35 (2002);
hep-th/0012238,
{\ Phys.~Lett.}~B \textbf{510}, 255 (2001).

\bibitem{dsrmost1} J.~Kowalski-Glikman:
hep-th/0102098,
Phys.~Lett.~A \textbf{286}, 391 (2001);
R.~Bruno, G.~Amelino-Camelia and J.~Kowalski-Glikman:
hep-th/0107039,
Phys.~Lett.~B \textbf{522}, 133 (2001).

\bibitem{kostenew} V.A.~Kostelecky, M.~Perry and R.~Potting,
Phys.Rev.Lett.84:4541-4544,2000, hep-th/9912243

\bibitem{gacpi} G.~Amelino-Camelia and S.-Y.~Pi:
hep-ph/9211211,
Phys.~Rev.~D \textbf{47}, 2356 (1993).

\bibitem{newlim} G.~Amelino-Camelia:
gr-qc/0212002.

\bibitem{rob} R.C.~Myers and M.~Pospelov:
hep-ph/0301124, Phys.~Rev.~Lett.~\textbf{90}, 211601 (2003).

\bibitem{billetal} S.D.~Biller {\it et al}:
{\ Phys.~Rev.~Lett.}~\textbf{83}, 2108 (1999).

\bibitem{glast}  J.P.~Norris, J.T.~Bonnell, G.F.~Marani and
J.D.~Scargle:
astro-ph/9912136;
A.~de Angelis:
astro-ph/0009271.

\bibitem{piranKARP}
T.~Piran,
astro-ph/0407462.

\bibitem{kifu} T.~Kifune:
{\ Astrophys.~J.~Lett.}~\textbf{518}, L21 (1999).

\bibitem{ita} R.~Aloisio, P.~Blasi, P.L.~Ghia and A.F.~Grillo:
Phys.~Rev.~D \textbf{62}, 053010 (2000).

\bibitem{aus} R.J.~Protheroe and H.~Meyer:
{\ Phys.~Lett.}~B \textbf{493}, 1 (2000).

\bibitem{gactp}  G. Amelino-Camelia and T. Piran:
astro-ph/0008107, {\ Phys.~Rev.}~D \textbf{64}, 036005 (2001).

\bibitem{jacoNATv1} T.~Jacobson, S.~Liberati and D.~Mattingly:
arXiv.org/abs/astro-ph/0212190v1


\bibitem{jaconature} T.~Jacobson, S.~Liberati and D.~Mattingly:
arXiv.org/abs/astro-ph/0212190v2,
Nature \textbf{424}, 1019 (2003).

\bibitem{tedreply} T.~Jacobson, S.~Liberati and D.~Mattingly:
gr-qc/0303001.

\bibitem{carrosync} S.~Carroll: Nature {\bf 424}, 1007 (2003).

\bibitem{nycksync} J.~Ellis, N.E.~Mavromatos and A.S.~Sakharov:
astro-ph/0308403.

\bibitem{tedsteck} T.A.~Jacobson, S.~Liberati, D.~Mattingly and F.W.~Stecker:
astro-ph/0309681.

\bibitem{jackson} J.D.~Jackson, \textit{Classical Electrodynamics},
3rd edn
(J.~Wiley \& Sons, New York 1999).

\bibitem{tedtwo} T.~Jacobson, S.~Liberati and D.~Mattingly:
hep-ph/0112207, Phys.~Rev.~D \textbf{66}, 081302 (2002); hep-ph/0209264.

\bibitem{gacpion} G. Amelino-Camelia: gr-qc/0107086,
{\ Phys.~Lett.}~B \textbf{528}, 181 (2002).

\bibitem{orfeupion} O.~Bertolami:
hep-ph/0301191.

\bibitem{nguhecr} Y.J.~Ng, D.S.~Lee, M.C.~Oh, and H.~van Dam:
Phys.~Lett.~B \textbf{507}, 236 (2001);
G. Amelino-Camelia, Y.J. Ng, and H. van Dam: gr-qc/0204077,
Astropart.~Phys.~\textbf{19}, 729 (2003).

\bibitem{agasa} M.~Takeda {\it et al}:
Phys.~Rev.~Lett.~\textbf{81}, 1163 (1998).

\bibitem{auger} J.~Blumer:
J.~Phys.~G \textbf{29}, 867 (2003).

\bibitem{bjpap} G.~Amelino-Camelia, J.D.~Bjorken and S.E.~Larsson,
hep-ph/9706530,
Phys.Rev.D56:6942-6956,1997

\bibitem{polonpap} G. Amelino-Camelia, ``Are we at the dawn
of quantum-gravity phenomenology?",
gr-qc/9910089, Lect.~Notes Phys.~541 (2000) 1;
 ``Quantum-gravity phenomenology: status and prospects",
gr-qc/0204051, Mod.~Phys.~Lett.~A17 (2002) 899.

\bibitem{dsrmost2} S.~Alexander and J.~Magueijo:
hep-th/0104093;
J.~Magueijo and L.~Smolin:
gr-qc/0207085, Phys.~Rev.~D \textbf{67}, 044017 (2003);
J.~Kowalski-Glikman and S.~Nowak:
hep-th/0304101,
Class.~Quant.~Grav.~\textbf{20}, 4799 (2003).

\bibitem{dsrphen} G.~Amelino-Camelia, J.~Kowalski-Glikman,
G.~Mandanici and A.~Procaccini:
gr-qc/0312124.

\end{thebibliography}
\end{document}